\documentclass[11pt,final,fleqn]{article}\usepackage[]{graphicx}\usepackage[]{color}

\usepackage{authblk}
\usepackage[T1]{fontenc}
\usepackage[margin=1in] { geometry }
\usepackage{amssymb,amsmath, bm}
\usepackage{verbatim}
\usepackage[latin1]{inputenc}
\usepackage{setspace}
\usepackage{enumitem}
\usepackage[hyphens,spaces,obeyspaces]{url}
\usepackage[labelfont={bf, footnotesize}, font = {footnotesize}]{caption}
\usepackage{latexsym}
\usepackage{graphicx}
\usepackage{marvosym}
\usepackage{pdflscape}
\usepackage{algorithm}

\usepackage{soul} 
\usepackage{color}
\usepackage{hyperref}
\usepackage{xcolor}
\hypersetup{
    colorlinks,
    linkcolor={blue!50!black},
    citecolor={blue!50!black},
    urlcolor={blue!80!black}
}
\hypersetup{pdfstartview={XYZ null null 1.00}} 

\usepackage{longtable}
\usepackage{booktabs, threeparttable}
\usepackage{threeparttablex}
\usepackage{tabularx}
\usepackage{dcolumn}
\newcolumntype{.}{D{.}{.}{-1}}
\newcolumntype{d}[1]{D{.}{.}{#1}}
\usepackage{multirow}
\usepackage[figuresright]{rotating}
\usepackage{pdflscape}
\usepackage{subcaption}
\usepackage{caption} 
\usepackage{grffile}
\usepackage{afterpage}
\usepackage{float}
\usepackage[section]{placeins}
\usepackage[export]{adjustbox}

\usepackage[compact]{titlesec}
\usepackage{etoolbox}
\BeforeBeginEnvironment{equation}{\begin{singlespace}}
\AfterEndEnvironment{equation}{\end{singlespace}\noindent\ignorespaces}
\BeforeBeginEnvironment{align}{\begin{singlespace}}
\AfterEndEnvironment{align}{\end{singlespace}\noindent\ignorespaces}

\newcommand{\pkg}[1]{{\fontseries{m}\fontseries{b}\selectfont #1}}

\def \medianLoghrTrtEcSe{0.148}
\def \hrMin{0.543}
\def \hrMax{1.586}
\def \muMedian{-0.098}
\def \muExpMedian{0.907}
\def \muExpLower{0.819}
\def \muExpUpper{1.007}
\def \sigmaMedian{0.114}
\def \sigmaLower{0.014}
\def \sigmaUpper{0.263}
\def \hrNewTrtIc{0.773}
\def \muExpMedianNoOutlier{0.876}
\def \muExpLowerNoOutlier{0.802}
\def \muExpUpperNoOutlier{0.957}
\def \sigmaMedianNoOutlier{0.061}
\def \sigmaLowerNoOutlier{0.005}
\def \sigmaUpperNoOutlier{0.168}



\title{A meta-analytic framework to adjust for bias in external control studies}

\author[1]{Devin Incerti}
\author[2]{Michael T Bretscher}
\author[1]{Ray Lin}
\author[3]{Chris Harbron}
\affil[1]{Genentech, Inc, South San Francisco, CA, USA}
\affil[2]{F. Hoffmann-La Roche Ltd, Basel, Switzerland}
\affil[3]{Roche Products, Welwyn Garden City, UK}
\date{\today}

\providecommand{\keywords}[1]
{
  \small	
  \textbf{\textit{Keywords---}} #1
}

\begin{document}
\maketitle

\begin{abstract}

	While randomized controlled trials (RCTs) are the gold standard for estimating treatment effects in medical research, there is increasing use of and interest in using real-world data for drug development. One such use case is the construction of external control arms for evaluation of efficacy in single-arm trials, particularly in cases where randomization is either infeasible or unethical. However, it is well known that treated patients in non-randomized studies may not be comparable to control patients---on either measured or unmeasured variables---and that the underlying population differences between the two groups may result in biased treatment effect estimates as well as increased variability in estimation. To address these challenges for analyses of time-to-event outcomes, we developed a meta-analytic framework that uses historical reference studies to adjust a log hazard ratio estimate in a new external control study for its additional bias and variability. The set of historical studies is formed by constructing external control arms for historical RCTs, and a meta-analysis compares the trial controls to the external control arms. Importantly, a prospective external control study can be performed independently of the meta-analysis using standard causal inference techniques for observational data. We illustrate our approach with a simulation study and an empirical example based on reference studies for advanced non-small cell lung cancer. In our empirical analysis, external control patients had lower survival than trial controls (hazard ratio: $\muExpMedian$), but our methodology is able to correct for this bias. An implementation of our approach is available in the \textsf{R} package \pkg{ecmeta}. 

\end{abstract}

\keywords{External controls, meta-analysis, bias, survival analysis, RCTs}

\section{Introduction}
Randomized controlled trials (RCTs) are considered the gold standard for estimation of treatment effects in medical research. Randomization ensures that no variables---measured or unmeasured---are systematically related to treatment assignment and that, consequently, the resulting treatment effect estimates are unbiased. However, there are potential barriers that limit the use of RCTs in certain settings: for example, a randomized control may be deemed unethical for disease areas with high unmet need or infeasible for targeted therapies with low expected enrollment.\cite{black1996we, simon2015role}. Thus, despite the scientific advantages of RCTs, single-arm trials are common during early drug development \cite{sambucini2015comparison} and have even been used for regulatory approval.\cite{hatswell2016regulatory}. 

At the same time, interest in using  real-world data (RWD) for drug development is increasing, as evidenced by the 21st Century Cures Act, which requires the U.S. Food and Drug Administration to evaluate the use of real-world evidence to support approval of new indications or to support post-approval study requirements. One such application of RWD is the construction of external control arms for single-arm trials so that efficacy can be assessed without a randomized control.\cite{burcu2020real, schmidli2020beyond, burger2021use} These estimates may not only be useful for regulators, but can guide internal decision making on drug development in biopharmaceutical companies as well.\cite{beyer2020multistate}

Unfortunately, it is well known that there are many potential sources of bias in non-randomized studies.\cite{hernan2020} In external control studies, potential biases are concerning because they may result in inflated type I error, which may be of particular importance for regulatory applications. When external controls are constructed using RWD, bias may be caused by systematic differences between trial and RWD patient populations that are difficult to control for. For instance, trial patients may tend to have better outcomes since they are more likely to be treated in specialized academic centers and given higher levels of attention. Furthermore, endpoints may be measured differently (e.g. progression endpoints in oncology) or assessed with different levels of quality in routine clinical practice than in clinical trials.  Finally, an important consideration is the extent to which outcomes for a given disease area can be explained by the observable data. If there are unobserved and highly prognostic confounders that vary across studies, then between study variability will be high and estimates from an external control study will be less certain.\cite{burger2021use} 

While external control analyses are relevant to a range of disease areas with different primary endpoints, we consider time-to-event outcomes here, although the proposed methodology could be readily adapted to other classes of endpoints. This is motivated, in part, by the oncology setting where overall survival (OS) is typically the primary endpoint in phase III studies. OS is attractive for external control studies because it is measured in the same way in trials and RWD, ensuring that differences between trial and external control patients are not due to measurement differences. A focus on oncology may also be warranted given that much of the external control work to date has focused on diseases with high unmet need such as oncology and rare diseases.\cite{feinberg2020use, wu2020use} Indeed, as precision medicine has become more common, the number of eligible patients in oncology trials has decreased and some oncology indications have paradoxically become ``rare diseases''.\cite{thorlund2020synthetic}

It is therefore of interest to develop methods for analysis of time-to-event outcomes that can adjust for the bias and the additional variability caused by external controls. 
In a prospective single-arm study, causal inference methods for observational data such as propensity score techniques are used to estimate the hazard ratio (HR) of the experimental treatment arm relative to the external control. In this paper, we propose a meta-analytic approach that leverages a set of historical reference studies to adjust HR estimates from such a study.  Each reference study includes the control arm of a clinical trial (i.e., the internal control) and an external control that is constructed based on the inclusion and exclusion criteria of the trial. To further increase comparability, propensity score models can be developed for each reference study to match the external control to the randomised treatment arm. A meta-analysis of the reference studies on the log HR scale is then performed to assess the compatibility of the internal and propensity score-matched external control arms. The treatment effects from the prospective single-arm study, are, in turn, adjusted based on the bias and between-study variability estimated from the meta-analysis. Importantly, the analysis for the single-arm study is conducted independently of the meta-analysis so that adjustment requires no modifications to the standard causal inference methods used in analyses with external controls. 

We use a simulation study to assess the operating characteristics of our methodology under different scenarios and then apply our method to a hypothetical single-arm study for treatment of advanced non-small cell lung cancer (aNSCLC), where we were able to use a set of 14 reference studies in the meta-analysis. Using a prespecified analysis plan that utilized propensity score methods, we estimated a mean bias (internal control vs. external control) on the log HR scale across the studies of $\muMedian$, or a HR of approximately $\muExpMedian$; that is, our analysis suggests that external control patients had shorter survival than trial controls, which would bias results in favor of the experimental treatment. The between study variability (on the log HR scale) was $\sigmaMedian$ as the HR varied from a low of $\hrMin$ to a high of $\hrMax$ across studies. An adjustment for the mean bias and between study variability would consequently widen uncertainty intervals and increase point estimates of the HR (i.e., decrease the treatment benefit). 

The remainder of this paper is organized as follows. \autoref{sec:methodology} introduces the meta-analytic framework. \autoref{sec:simulations} presents results from the simulation study and \autoref{sec:example} reports results from the aNSCLC example. \autoref{sec:discussion} puts our methodology in context,  discusses challenges and key considerations for implementation, and notes other potential applications. Finally, \autoref{sec:conclusion} concludes. 

\section{Methodology} \label{sec:methodology}

\subsection{Model formulation}
We consider a set of $n+1$ studies, $n$ reference studies and a new study (i.e., a prospective single-arm trial), each of which may be considered to have 3 arms:

\begin{itemize}
\item $TRT$: the randomized experimental treatment arm
\item $IC$: the randomized internal control arm
\item $EC$: the external control arm
\end{itemize}

\begin{figure}[t!]
\centering
\begin{subfigure}[b]{\textwidth}
\includegraphics[max size={.9\textwidth}{\textheight}]{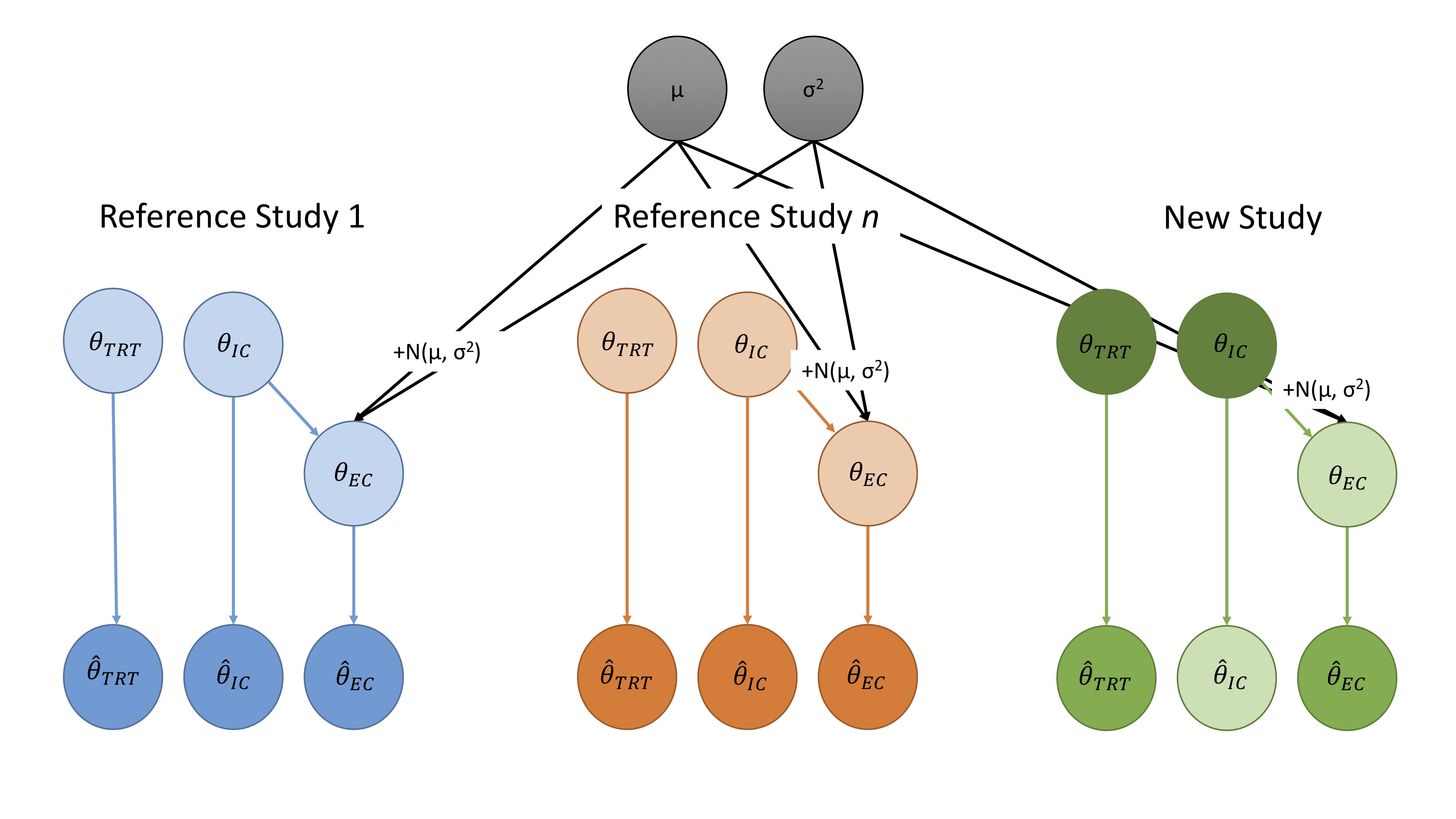} 
\caption{Absolute outcomes} \label{fig:dag-absolute}
\end{subfigure}
\begin{subfigure}[b]{\textwidth}
\includegraphics[max size={.9\textwidth}{\textheight}]{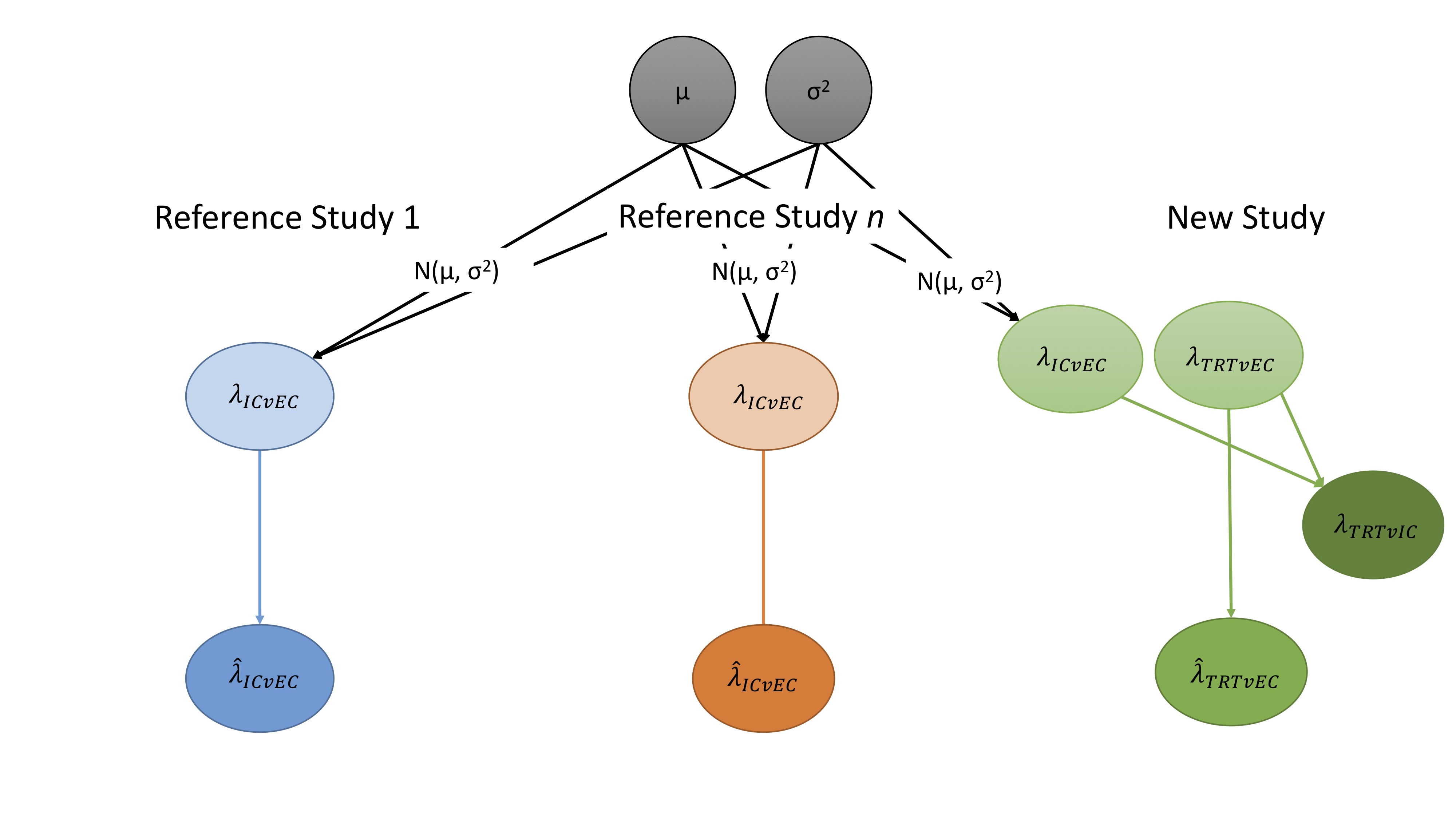} 
\caption{Hazard ratios} \label{fig:dag-relative}
\end{subfigure}
\caption{Directed acyclic graph describing the data generating process for survival outcomes. $\theta$ contains parameters describing the generating process for survival outcomes and $\lambda$ represents a log hazard ratio. The ``hat'' distinguishes an estimated parameter from the true value of a parameter. $\mu$ and $\sigma$ model the degree of similarity between the internal and external controls. The light colored nodes are not observable while the moderately dark colored nodes are observable. The darkest colored nodes reflect the quantities of interest: our aim is to estimate treatment effects by comparing the experimental treatment arm to a hypothetical internal control arm, which is defined in terms of log HRs as $\lambda_{TRTvIC}$.}
\label{fig:dag}
\end{figure}

\autoref{fig:dag-absolute} represents the data generating process for survival outcomes in terms of a directed acyclic graph. We let $\theta$ be the ``true'' values of the parameters that generate survival outcomes (e.g., $\theta$ might represent the rate parameter in an exponential survival model or the shape and scale parameters in a Weibull model). The estimates of the parameters in the study datasets are denoted by $\hat{\theta}$. The light colored nodes are not observable while the moderately dark colored nodes are observable. Our aim is to estimate treatment effects by comparing survival outcomes generated by the darkest nodes, $\theta_{TRT}$ to $\theta_{IC}$, in the new study. The challenge is of course that we have no randomized control so $\hat{\theta}_{IC}$ is not observable in the new study and a standard approach that compares $\hat{\theta}_{TRT}$  to $\hat{\theta}_{EC}$  is likely to result in bias because the external control arm is not randomly assigned. As such, we model the degree of similarity between the outcomes in the internal and external control arms using a normal distribution with parameters for the mean, $\mu$, and variance, $\sigma^2$. 

It is of note that neither $\hat{\theta}_{TRT}$ nor $\theta_{TRT}$ from the references studies impact survival outcomes in the new study. We can consequently disregard the outcomes for the treatment arms from the reference studies and reformulate our problem in terms of HRs as shown in \autoref{fig:dag-relative}, where $\lambda$ is the true log HR and $\hat{\lambda}$ is an estimate of the log HR in the study sample. In other words, we can focus solely on comparisons between the internal and external controls in the reference studies.

In what follows, we describe a meta-analytic model for estimation of log HRs for comparisons between the internal and external controls, $\lambda_{ICvEC}$, and an approach for generating the posterior distribution of our quantity of interest, $\lambda_{TRTvIC}$. While our primary analysis uses Bayesian techniques for parameter estimation, quantification of uncertainty, and prediction, \autoref{appendix:ml} details an alternative simulation based approach based on maximum likelihood estimation. We discuss the pros and cons of each as well as considerations for priors in the Bayesian model in \autoref{sec:discussion}. The methodology can be implemented using the publicly available \textsf{R} package \pkg{ecmeta}.

\subsection{Meta-analytic model} \label{subsec:meta-analytic}
A comparison between the internal and external controls in terms of the log HR can be modeled as,

\begin{equation}
\hat\lambda_{ICvEC,j} \sim N (\lambda_{ICvEC,j}, S_j^2),
\end{equation}

where $j$ indexes a study and $S_j^2$ is the sampling variance. We assume that the true log HR is distributed with mean $\mu$ and variance $\sigma^2$, 

\begin{equation}
\lambda_{ICvEC,j} \sim N ( \mu , \sigma^2),
\end{equation}

so that $\mu$ is a measure of the average bias in the external control arms across studies and $\sigma^2$ is a measure of between study variability. It follows that the estimated log HR in study $j$ is distributed as,

\begin{equation}
\hat\lambda_{ICvEC,j} \sim N ( \mu , \sigma^2 + S_j^2).
\end{equation}

When a Bayesian approach is used for estimation, priors are placed on $\mu$ and $\sigma$. We use a normal distribution for $\mu$ with mean $0$ and variance large enough so that the prior is uninformative. The prior for $\sigma$ is arguably a more important consideration since it usually has a larger influence on the posterior distribution, particularly when the number of reference studies is small. In our software implementation, we allow for three uninformative priors that have been suggested in the literature including uniform \cite{gelman2013bayesian} and half-t \cite{gelman2006prior, polson2012half} distributions for $\sigma$ and a $\text{Gamma}(0.001, 0.001)$ distribution \cite{spiegelhalter2004bayesian} for $1/\sigma^2$.

\subsection{Prediction for a new study} \label{subsec:prediction}
In a RCT, the treatment effect is estimated in an unbiased manner by comparing the experimental treatment arm to the internal control arm. Since there is no internal control arm in a single-arm trial, we adjust the HR from the single-arm study to obtain a HR that we would have obtained had an internal control existed. Such an adjustment is made by leveraging the following relationship between log HRs,

\begin{align} \label{eqn:hr-relationship}
\lambda_{TRTvIC}^{new} = \lambda_{TRTvEC}^{new} - \lambda_{ICvEC}^{new}.
\end{align}

Note that this relationship will hold exactly for the true log HRs under the assumption of proportional hazards. For estimated log HRs (i.e., $\hat{\lambda}$), the relationship is only an approximation and will have a small error attached to it. 

A prediction for $\lambda_{TRTvIC}^{new} $ in a new study then proceeds in three steps. The first step draws posterior samples of $ \lambda_{TRTvEC}^{new}$ after using standard causal inference methods for observational data (such as those from a propensity score adjusted Cox regression) to estimate the log HR, $\hat{\lambda}_{TRTvEC}^{new}$, and its variance, $V^{new}$. Conveniently, no modifications to standard analyses with external controls are required.  A model for the sampling distribution for $\lambda_{TRTvEC}^{new}$ (using $\hat{\lambda}_{TRTvEC}^{new}$ as data and treating $V^{new}$ as fixed)  is given by,

\begin{align}
\hat{\lambda}_{TRTvEC}^{new} &\sim N(\lambda_{TRTvEC}^{new}, V^{new}),
\end{align}

where a normal prior is used for $\lambda_{TRTvEC}^{new}$ with mean $0$ and large variance. The second step samples $\lambda_{ECvIC}^{new}$ using posterior draws of $\mu$ and $\sigma$ from the meta-analytic model,

\begin{align}
\lambda_{ICvEC}^{new} ~ \sim N ( \mu , \sigma^2).
\end{align}

Finally, in the third step, the posterior draws of $\lambda_{TRTvEC}^{new}$ and $\lambda_{ICvEC}^{new}$ are combined to obtain a posterior sample of $\lambda_{TRTvIC}^{new}$ using \autoref{eqn:hr-relationship}.

\section{Simulations} \label{sec:simulations}

\subsection{Setup and scenarios}
Six scenarios were simulated as described in \autoref{tbl:sim-scenarios}. For each scenario, parameters for 10,000 studies were generated as a function of median survival time and the number of events in each of the three arms. The log of median survival and the log of the number of events across studies were modeled using normal distributions centered at the medians displayed in the table and with variability determined by the coefficient of variation; that is, median survival times and the number of events were drawn from lognormal distributions with location parameters equal to the medians and the scale parameters equal to the coefficients of variation.

\begin{table}
\scriptsize
\begin{center}
\begin{threeparttable}
\caption{Simulation scenarios} \label{tbl:sim-scenarios}
\begin{tabular}{lm{6cm}llllll}
\hline
\multicolumn{2}{l}{} & \multicolumn{3}{c}{Median survival time} & \multicolumn{3}{c}{Number of events} \\
\multicolumn{2}{l}{} & \multicolumn{3}{c}{[Median (CV)$^\dagger$]} & \multicolumn{3}{c}{[Median (CV)$^\dagger$]} \\
\cmidrule(r){3-5} \cmidrule(r){6-8}
\multicolumn{1}{l}{} & \multicolumn{1}{l}{Description} & \multicolumn{1}{l}{Treatment} &  \multicolumn{1}{l}{Internal}  & \multicolumn{1}{l}{External} & \multicolumn{1}{l}{Treatment} &  \multicolumn{1}{l}{Internal}  & \multicolumn{1}{l}{External} \\
\multicolumn{3}{l}{} & \multicolumn{1}{l}{control} &  \multicolumn{1}{l}{control} & \multicolumn{1}{l}{} & \multicolumn{1}{l}{control} &  \multicolumn{1}{l}{control}\\
\hline
S1 & 
There is no between study heterogeneity (i.e., each study has the same median survival) and the number of events does not vary between studies. & 
24 (0\%) & 15 (0\%) & 12 (0\%) & 100 (0\%) & 70 (0\%) & 50 (0\%)  \\
\hline
S2 &
There is no between study variability and the number of events in each arm varies between studies. & 
24 (0\%) & 24 (0\%) & 18 (0\%) & 250 (20\%) & 250 (20\%) & 250 (20\%) \\
\hline
S3 &
There is independent between-study variability in all three arms and variability in the number of events in each arm between studies. & 
24 (40\%) & 24 (20\%) & 18 (20\%) & 250 (20\%) & 250 (20\%) & 250 (20\%) \\
\hline
S4 &
A null-hypothesis scenario where the true randomized hazard ratio is fixed at 1 for all studies and the external control is biased with the magnitude of bias varying between studies. The number of events in the randomized study and external control arm vary between studies and the two randomized arms are of equal size (i.e., a 1:1 randomization ratio). & 
\multicolumn{2}{c}{24 (20\%) - HR=1$^+$} &
18 (20\%) &
\multicolumn{2}{c}{150 (20\%)*} &
250 (20\%) \\
\hline
S5 &
An alternative-hypothesis scenario where the true randomized hazard ratio is fixed at 0.5 for all studies and the external control is biased with the magnitude of bias varying between studies. The number of events in the randomized study and external control arm vary between studies and the two randomized arms are of equal size (i.e., a 1:1 randomization ratio). &
\multicolumn{2}{c}{48/24 (20\%) - HR=0.5$^+$} &
18 (20\%) &
\multicolumn{2}{c}{150 (20\%)*} &
250 (20\%) \\
\hline
S6 &
An alternative-hypothesis scenario where the randomized hazard ratio varies between studies and the external control is biased with the magnitude of bias varying between studies (this may simulate scenarios where the reference set of studies are drawn from studies examining a class of multiple different compounds). The number of events in the randomized study and external control arm vary between studies and the two randomized arms are of equal size (i.e., a 1:1 randomization ratio). &
35 (40\%) &
24 (20\%) &
18 (20\%) &
\multicolumn{2}{c}{250 (20\%)*} &
250 (20\%) \\
\hline
\end{tabular}
\scriptsize 
$^\dagger$CV = coefficient of variation. Median survival times and the number of events were drawn from lognormal distributions with location parameters equal to the medians in the table and scale parameters equal to the coefficient of variation. \\
*The sizes of the two RCT arms were fixed to be equal for each study assuming a 1:1 randomization ratio. \\
$^+$The true hazard ratio between the two RCT arms was fixed.
\end{threeparttable}
\end{center}
\end{table}

Parameters were always simulated independently for the external control arms. Parameters for each arm of the RCT were simulated in different ways depending on the scenario. In some scenarios the parameters of the two randomized arms were simulated independently of each other (S1, S2, S3) and in other scenarios they were set according to a fixed HR or a fixed 1:1 randomization ratio (S4, S5, S6).

Patient level data was simulated separately for each arm and study using exponential distributions where the rate parameter was set according to the median survival times. A Cox model was then used to estimate three log HRs---$\hat{\lambda}_{TRTvIC}$, $\hat{\lambda}_{TRTvEC}$, and $\hat{\lambda}_{ICvEC}$---for each study.  Note that we assumed, for simplicity, that there was no censoring, meaning that the sample size was equal to the number of events. Given that variability in the log HRs is driven by the number of events, this assumption should not be influential. 

To evaluate the meta-analytic methodology and assess the impact of varying the number of references studies, the 10,000 simulated studies were split into $M=10,000/(n+1)$ replications where $n$ was the number of reference studies, which we varied from 4 to 9. In each replication, $n$ studies were used to estimate the meta-analytic model and the remaining study was used as the new single-arm study for which a prediction of $\lambda^{new}_{TRTvIC}$ was made. 

In \autoref{sec:results}, we report estimates from a Bayesian model and consequently specified priors for the hyperparameters. We used a diffuse $N(0, 100)$ prior for $\mu$ and following the advice of Gelman \cite{gelman2006prior}, a half-Cauchy (a half-t with 1 degree of freedom) prior for $\sigma$ with location $0$ and a large scale parameter ($A = 25$). In our discussion (\autoref{sec:discussion}) we consider alternative approaches including uniform and inverse gamma priors as well as estimation via maximum likelihood. Plots summarizing evaluations of these alternative approaches are provided in the Appendix (\autoref{appendix:figs}).

\subsection{Results} \label{sec:results}
\begin{figure}[t!]
\centering
\includegraphics[max size={.8\textwidth}{\textheight}]{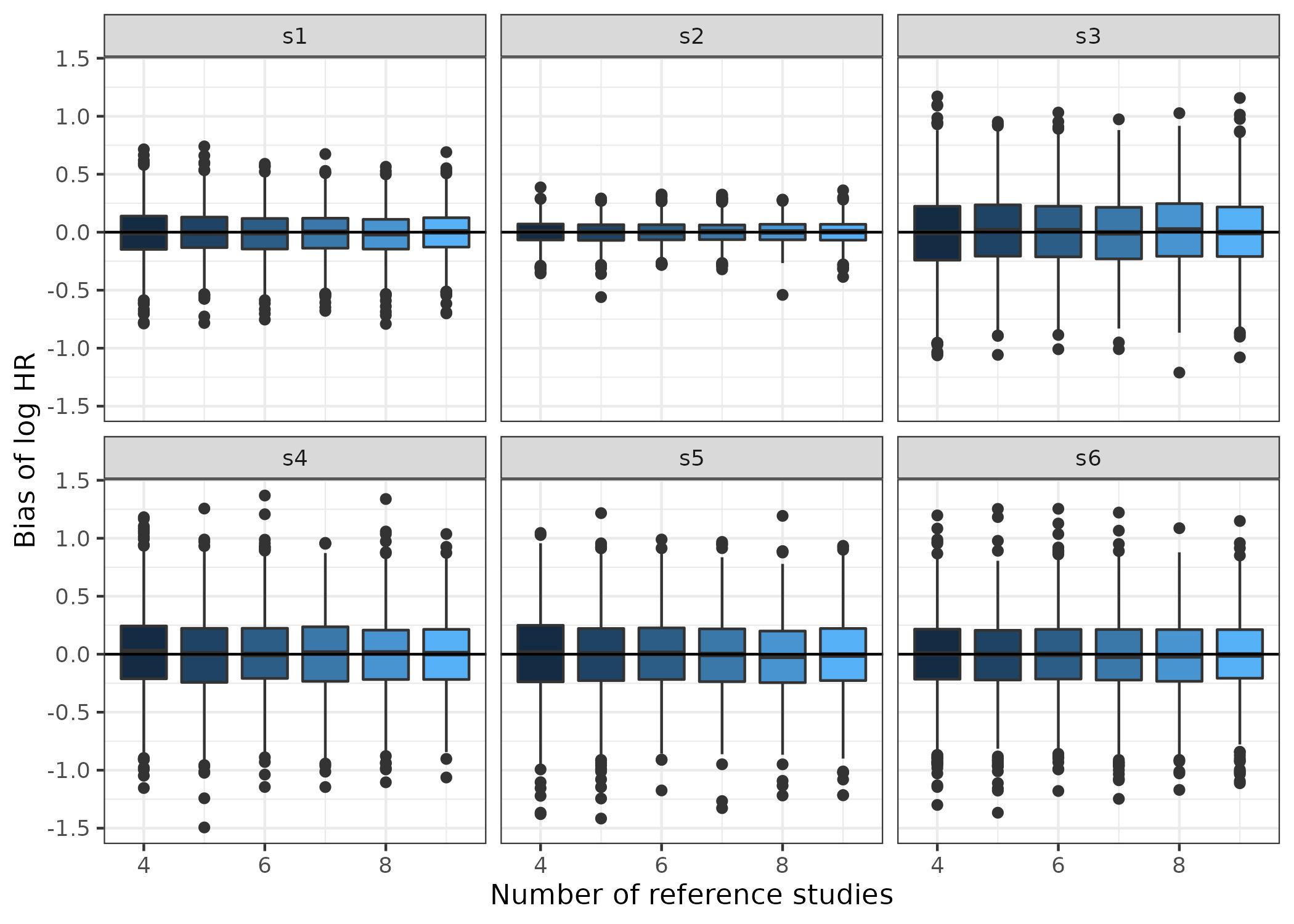} 
\caption{Bias of the adjusted estimates of the log hazard ratio as a function of the simulation scenario and the number of reference studies. The meta-analysis was performed using a Bayesian model with a half-Cauchy prior for $\sigma$.}
\label{fig:bias-cauchy}
\end{figure}

\autoref{fig:bias-cauchy} plots estimates of bias by scenario and the number of reference studies used to fit the meta-analytic model. The point estimate for the log HR, $\hat{\lambda}^{new}_{TRTvIC}$, was estimated using the median of the posterior distribution and bias was computed for the $m$th replication as $\hat{\lambda}^{new}_{TRTvIC,m} - \lambda^{new}_{TRTvIC,m}$ where $\lambda^{new}_{TRTvIC,m}$ is the true value of the log HR from the simulation. The boxplot characterizes the distribution of bias across the simulation replications. There is, in general, no evidence of bias as the median value of bias across the replications was approximately 0 in each scenario. Similarly, there is no evidence that the number of reference studies has a discernible impact on bias. 

\begin{figure}[t!]
\centering
\includegraphics[max size={.8\textwidth}{\textheight}]{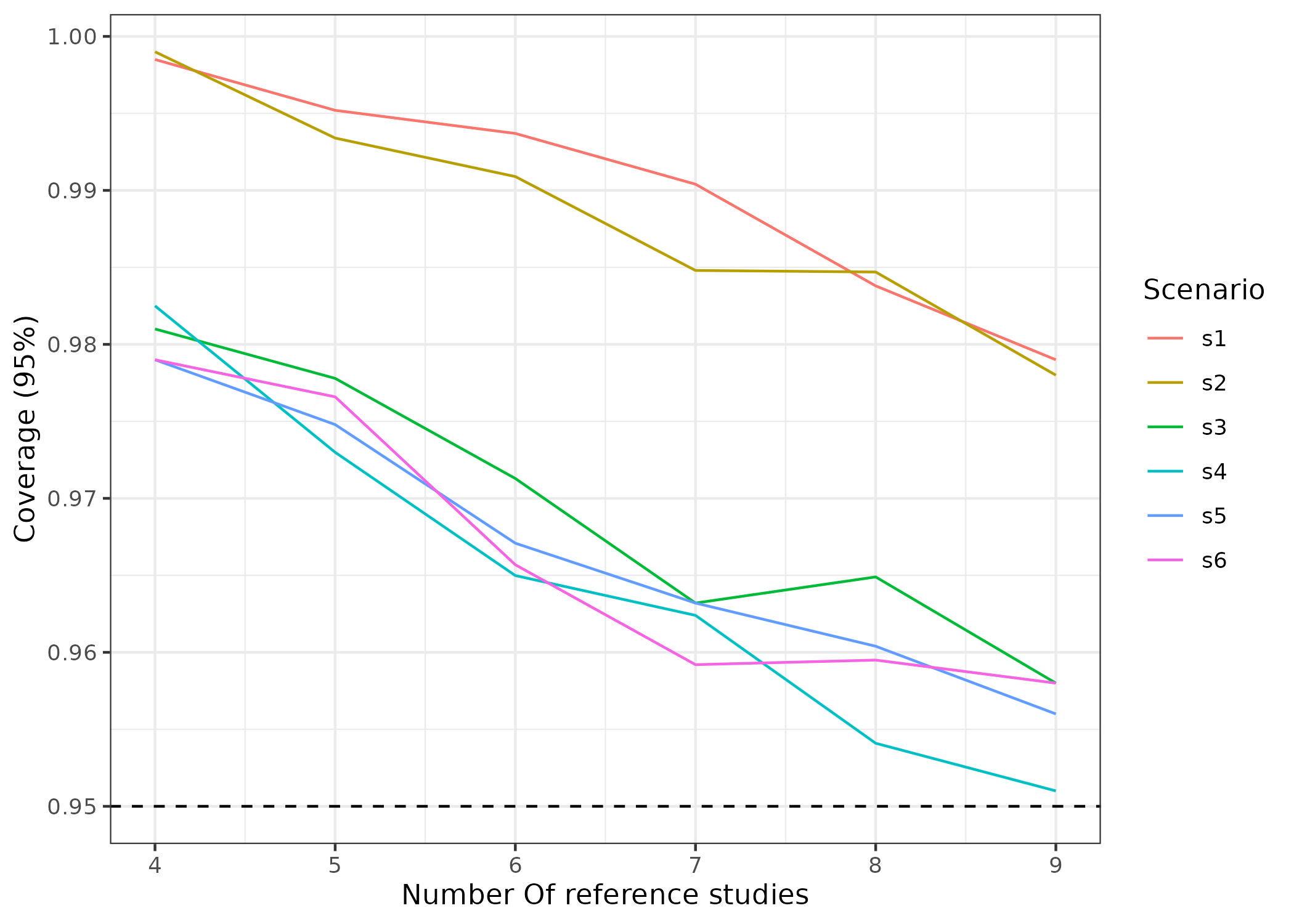} 
\caption{Coverage of 95\% confidence intervals of the adjusted estimates of the log hazard ratio as a function of the simulation scenario and number of reference studies. The meta-analysis was performed a using Bayesian model with a half-Cauchy prior for $\sigma$.}
\label{fig:coverage-cauchy}
\end{figure}

The coverage of nominal 95\% confidence intervals are plotted in \autoref{fig:coverage-cauchy}. The results are conservative with some overcoverage since the 95\% credible intervals (CrIs) from the model contain the true value of the log HR, $\lambda^{new}_{TRTvIC, m}$, more often than the nominal 95\%. The extent of overcoverage is decreasing in the number of reference studies, which suggests that a larger set of reference studies can improve the precision of the estimates.

\begin{figure}[t!]
\centering
\includegraphics[max size={.8\textwidth}{\textheight}]{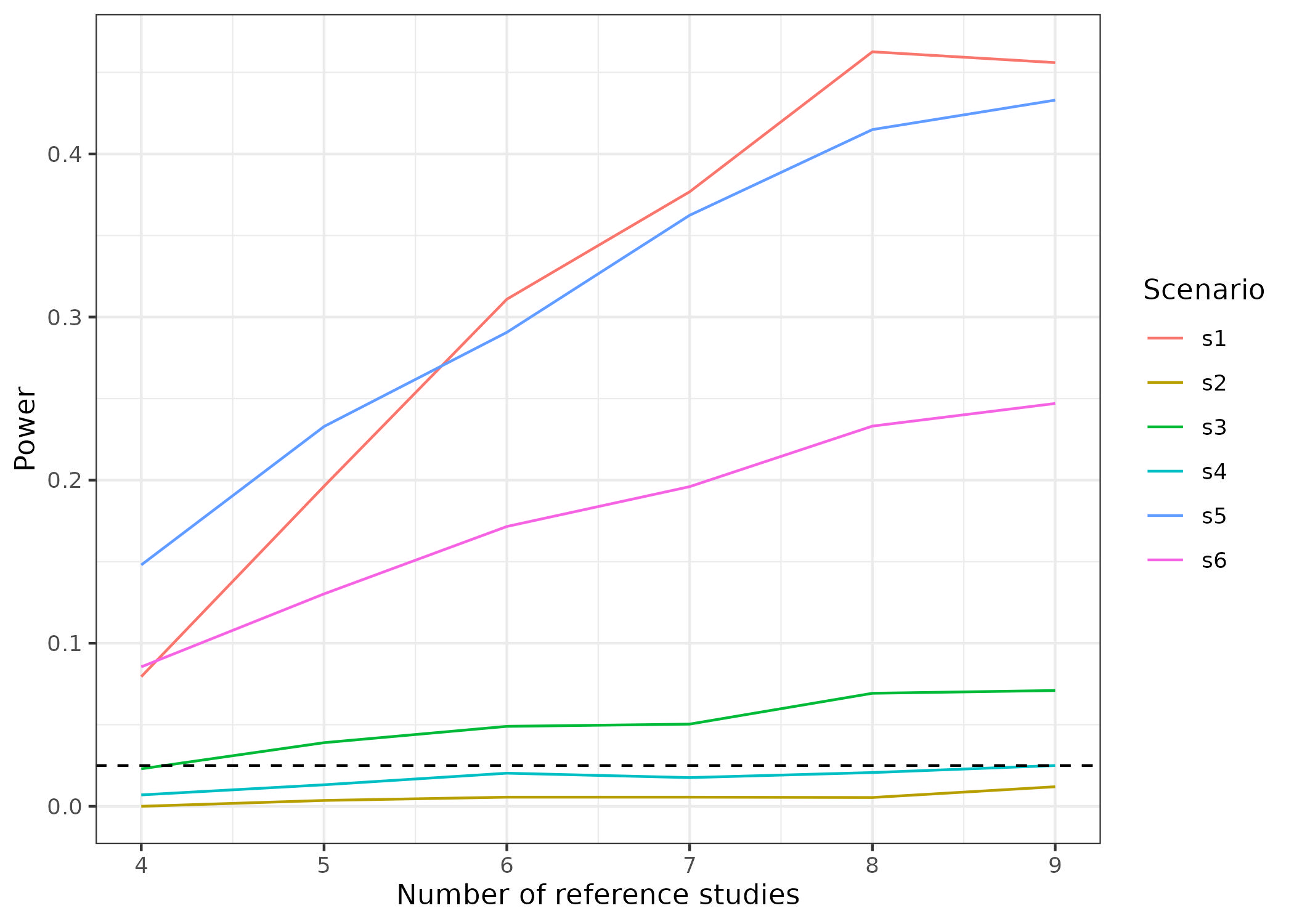} 
\caption{Power of adjusted estimates of the log hazard ratio by scenario and the number of reference studies. Power (S1, S3, S5, S6) and type I error (S2, S4) were computed from a one-sided test. The dotted horizontal line represent the expected type I error of 0.025 under the null. The meta-analysis was performed a using Bayesian model with a half-Cauchy prior for $\sigma$.}
\label{fig:power-cauchy}
\end{figure}

One consequence of our methodology is that uncertainty intervals will increase after adjustment. We assessed the implications of this on the power of the Bayesian model in \autoref{fig:power-cauchy}. Power and type I error were computed from a one-sided test where a treatment effect was deemed significant if the upper limit of the 95\% CrI was less than zero. In null-hypothesis scenarios---S2 and S4---where the true log HR was 1, the type I error was below $0.025$ (represented by the dotted line), suggesting that the methodology was able to prevent type I error inflation. In the remaining scenarios, power was generally increasing in the number of reference studies. Increasing the number of reference studies therefore seems beneficial since it reduces uncertainty (\autoref{fig:coverage-cauchy}) and increases power.

\section{ Example: Advanced non-small cell lung cancer} \label{sec:example}

\subsection{Reference studies}
We constructed a set of 14 reference studies using phase II and phase III RCTs for treatment of patients with  aNSCLC. To ensure that patient-level data was available, we restricted trials to those conducted by Roche (which we had access to). The set of trials is the same as in a study by Carrigan et al.\cite{carrigan2020using}, but includes three additional trials: NCT02367781, NCT02367794, and NCT02657434. In some cases (NCT01496742 and NCT01493843), multiple treatment arms (with corresponding control arms) were available for a single trial. We treated those as separate studies and  refer to each unique treatment arm and control arm pair as a unique reference study in the remainder of this article. The full list of reference studies is shown in \autoref{tbl:reference-studies}.

A unique external control cohort was constructed for each of the internal control arms from RWD. To increase comparability between the internal and external controls, the inclusion and exclusion criteria from the trials were applied to the external control cohorts. Following Carrigan et al., we used the Flatiron health database---which provides retrospective, longitudinal de-identified clinical information derived from electronic health records---as our data source. The data originated from academic and community cancer clinics across the United States and contains both structured and unstructured (via technology-enabled abstraction) elements.\cite{birnbaum2020model, ma2020comparison} The data are de-identified and subject to obligations to prevent re-identification and protect patient confidentiality. Institutional review board approval of the study was obtained prior to study conduct, and included a waiver of informed consent.

\begin{table}
\scriptsize
\begin{center}
\begin{threeparttable}
\caption{Reference studies for meta-analysis} \label{tbl:reference-studies}
\begin{tabular}{llp{4cm}p{4cm}lll}
\hline
\multicolumn{2}{l}{} & \multicolumn{2}{c}{Comparison} & \multicolumn{3}{c}{Sample size} \\
\cmidrule(r){3-4} \cmidrule(r){5-7}
\multicolumn{1}{l}{} & \multicolumn{1}{l}{Clinical trial} & \multicolumn{1}{l}{Treatment} & \multicolumn{1}{l}{Control} & \multicolumn{1}{l}{Treatment} & \multicolumn{1}{l}{Internal}  & \multicolumn{1}{l}{External} \\
\multicolumn{5}{l}{} & \multicolumn{1}{l}{control} &  \multicolumn{1}{l}{control}\\
\hline
  \hline
  1 & NCT02008227 & Atezolizumab & Docetaxel & 425 & 425 & 526 \\ 
    2 & NCT01903993 & Atezolizumab & Docetaxel & 144 & 143 & 476 \\ 
    3 & NCT02366143 & Atezolizumab + Bevacizumab + Carboplatin & Bevacizumab + Carboplatin & 356 & 336 & 602 \\ 
    4 & NCT01351415 & Bevacizumab + SOC & SOC & 245 & 240 & 381 \\ 
    5 & NCT01519804 & MetMAb + Platinum + Paclitaxel & Placebo + Platinum + Paclitaxel &  55 &  54 & 1945 \\ 
    6 & NCT01496742 & MetMAb + Bevacizumab + Platinum + Paclitaxel & Placebo + Bevacizumab + Platinum &  69 &  70 & 956 \\ 
    7 & NCT01496742 & MetMAb + Platinum + Pemetrexed & Placebo + Platinum + Pemetrexed &  59 &  61 & 3244 \\ 
    8 & NCT01366131 & MEGF0444A + Bevacizumab + Carboplatin + Paclitaxel & Placebo + Bevacizumab + Carboplatin + Paclitaxel &  52 &  52 & 963 \\ 
    9 & NCT01493843 & Pictilisib (340 mg) + Carboplatin + Paclitaxel & Placebo (340 mg) + Carboplatin + Paclitaxel & 126 & 125 & 1274 \\ 
   10 & NCT01493843 & Pictilisib (340 mg) + Carboplatin + Paclitaxel + Bevacizumab & Placebo (340 mg) + Carboplatin + Paclitaxel + Bevacizumab &  79 &  79 & 953 \\ 
   11 & NCT01493843 & Pictilisib (260 mg) + Carboplatin + Paclitaxel + Bevacizumab & Placebo (260 mg) + Carboplatin + Paclitaxel + Bevacizumab &  62 &  30 & 953 \\ 
   12 & NCT02367781 & Atezolizumab + Nab-Paclitaxel + Carboplatin & Nab-Paclitaxel + Carboplatin & 451 & 228 &  87 \\ 
   13 & NCT02367794 & Atezolizumab + Nab-Paclitaxel + Carboplatin & Nab-Paclitaxel + Carboplatin & 343 & 340 & 497 \\ 
   14 & NCT02657434 & Atezolizumab + Carboplatin or Cisplatin + Pemetrexed & Carboplatin or Cisplatin + Pemetrexed & 292 & 286 & 1536 \\ 
   \hline

\hline
\end{tabular}
\scriptsize Note: All external control datasetets were built using the Flatiron Health database. Sample sizes in the external control datasets were based on the number of patients remaining in the data after applying the trial inclusion and exclusion criteria.  
\end{threeparttable}
\end{center}
\end{table}

\subsection{Propensity score methodology} \label{subsec:propensity-score}
We used propensity score methods to estimate log HRs (internal versus external control), $\hat{\lambda}_{ICvEC,j}$, and standard errors, $S_j$, in each study, $j$. We considered the estimand of interest to be the average treatment effect on the treated (ATT), which is the log HR among the trial participants. The time-to-event outcome was overall survival (OS).

To mitigate the risk of underestimating bias, we prespecified our analysis. Our primary analysis used inverse probability of treatment weights (IPTW) permitting estimation of the ATT (i.e., IPTW-ATT weights) so that internal controls received a weight of 1 and external controls received a weight of $e/(1-e)$ where $e$ is the propensity score.\cite{austin2015moving} To reduce the influence of extreme weights, external control patients with propensity scores below the 1st percentile and above the 99th percentile were removed. 

Propensity scores were modeled using logistic regression with the same covariates used by Carrigan et al.\cite{carrigan2020using}---age, sex, race (White, Black, other), histology (non-squamous, squamous), smoking status (current/former, never), disease stage at initial diagnosis (Advanced - IIIB/IV, Early - IIIA or below), and time from initial diagnosis until either the start of treatment (RWD) or randomization (trial data). Nonlinear functions of the continuous covariates---age and time from initial diagnosis---were considered, but did not improve covariate balance so simpler linear terms were employed.

HRs were estimated using IPTW-ATT weighted Cox proportional hazards models. The models were fit without covariate adjustment (i.e., they only included a single covariate for treatment assignment) to facilitate estimation of a marginal HR,\cite{daniel2021making} which is typically the estimand of interest from a RCT. Confidence intervals were computed using a robust sandwich-type variance estimator.\cite{austin2014use}. 

Missing covariates were imputed using multiple imputation with the \texttt{aregImpute} function from the \textsf{R}  package \pkg{Hmisc}.\cite{Hmisc2021} The data source (trial or RWD) was interacted with the covariates so that there was effectively a separate imputation model for the trial and external control patients. Given recommendations from recent simulation studies, we estimated log HRs separately in each imputed dataset and then estimated pooled log HRs and confidence intervals using Rubin's rules.\cite{leyrat2019propensity,granger2019avoiding}

Additional details and sensitivity analyses are described in \autoref{appendix:ps}.

\subsection{Estimation of meta-analytic model}
\begin{figure}[h]
\centering
\includegraphics[max size={.8\textwidth}{\textheight}]{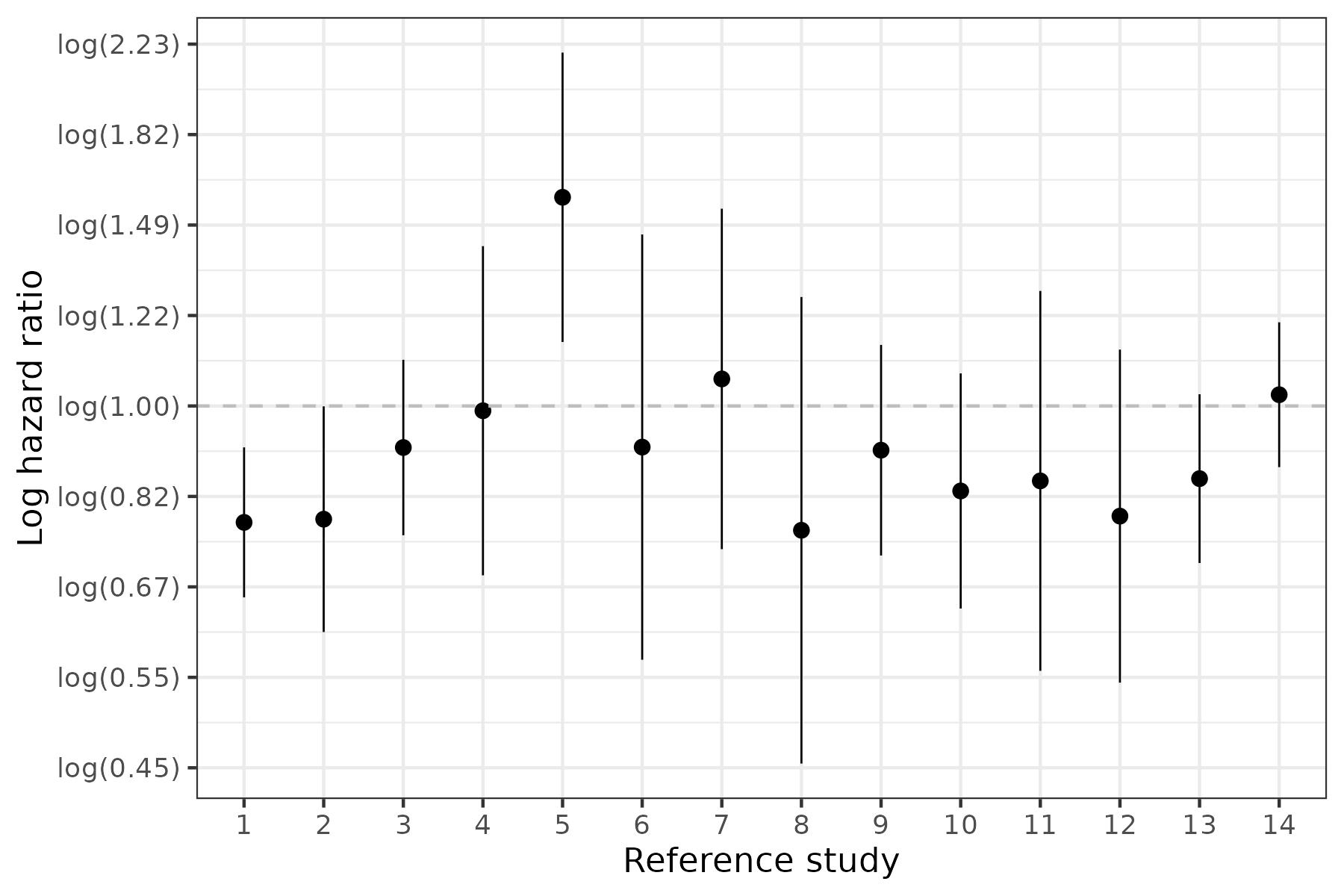} 
\caption{Log hazard ratios (HRs) for comparisons between internal and external control arms, by reference study. Error bars represent 95\% confidence intervals. Survival times are, on average, shorter for the external controls if the $\text{HR} < 1$ and longer if the $\text{HR} > 1$.}
\label{fig:loghr-ic-ec}
\end{figure}

The IPTW-ATT weighted estimates of the log HRs and 95\% confidence intervals are shown in \autoref{fig:loghr-ic-ec}. The HRs tend to be below 1, implying that survival is, on average, shorter among Flatiron patients than the internal controls. The implication is that a standard analysis that compares an experimental treatment arm to an external control in aNSCLC will tend to be biased in favor of the treatment.

A formal assessment of bias and between study variability was performed using the meta-analysis model described in \autoref{subsec:meta-analytic}. We estimated the model using the same Bayesian approach employed in the simulations described in \autoref{sec:simulations}. In particular, we used a diffuse $N(0, 100)$ prior for $\mu$ and a half-Cauchy prior for $\sigma$ with location $0$ and scale parameter $A = 25$.

The estimates were consistent with  \autoref{fig:loghr-ic-ec}: the posterior median for the bias, $\mu$, was $\log(\muExpMedian)$ with a 95\% CrI ranging from a lower limit (2.5th percentile) of $\log(\muExpLower)$ to an upper limit (97.5th percentile) of  $\log(\muExpUpper)$, again implying that the external controls tended to have shorter survival. The estimates for $\sigma$ (posterior median: $\sigmaMedian$; 95\% CrI: $\sigmaLower$ to $\sigmaUpper$) suggest that there is significant between study variability, which is not unexpected for a scale hyperparameter in a hierarchical model with a relatively small number of observations. A non-negligible fraction of the between study variability was caused by reference study 5, which appears to be a significant outlier: after removal, $\sigma$ decreased considerably (posterior median: $\sigmaMedianNoOutlier$; 95\% CrI: $\sigmaLowerNoOutlier$ to $\sigmaUpperNoOutlier$) and $\mu$ declined slightly (posterior median: $\log(\muExpMedianNoOutlier)$; 95\% CrI: $\log(\muExpLowerNoOutlier)$ to $\log(\muExpUpperNoOutlier)$).

\subsection{Hypothetical analysis for a new study}
To illustrate our approach, we considered a hypothetical single-arm trial with an external control arm in which the estimated HR, $\exp(\hat{\lambda}_{TRTvEC,j})$, was 0.70. To generate reasonable estimates of the standard error, we used the same propensity score methods described in \autoref{subsec:propensity-score} to compare the experimental treatment arms to the external controls and estimate log hazard ratios for each reference study $j$. We then set the standard error of the log hazard ratio equal to the median of the standard errors across the reference studies, $\medianLoghrTrtEcSe$.

\begin{figure}[h]
\centering
\includegraphics[max size={.9\textwidth}{\textheight}]{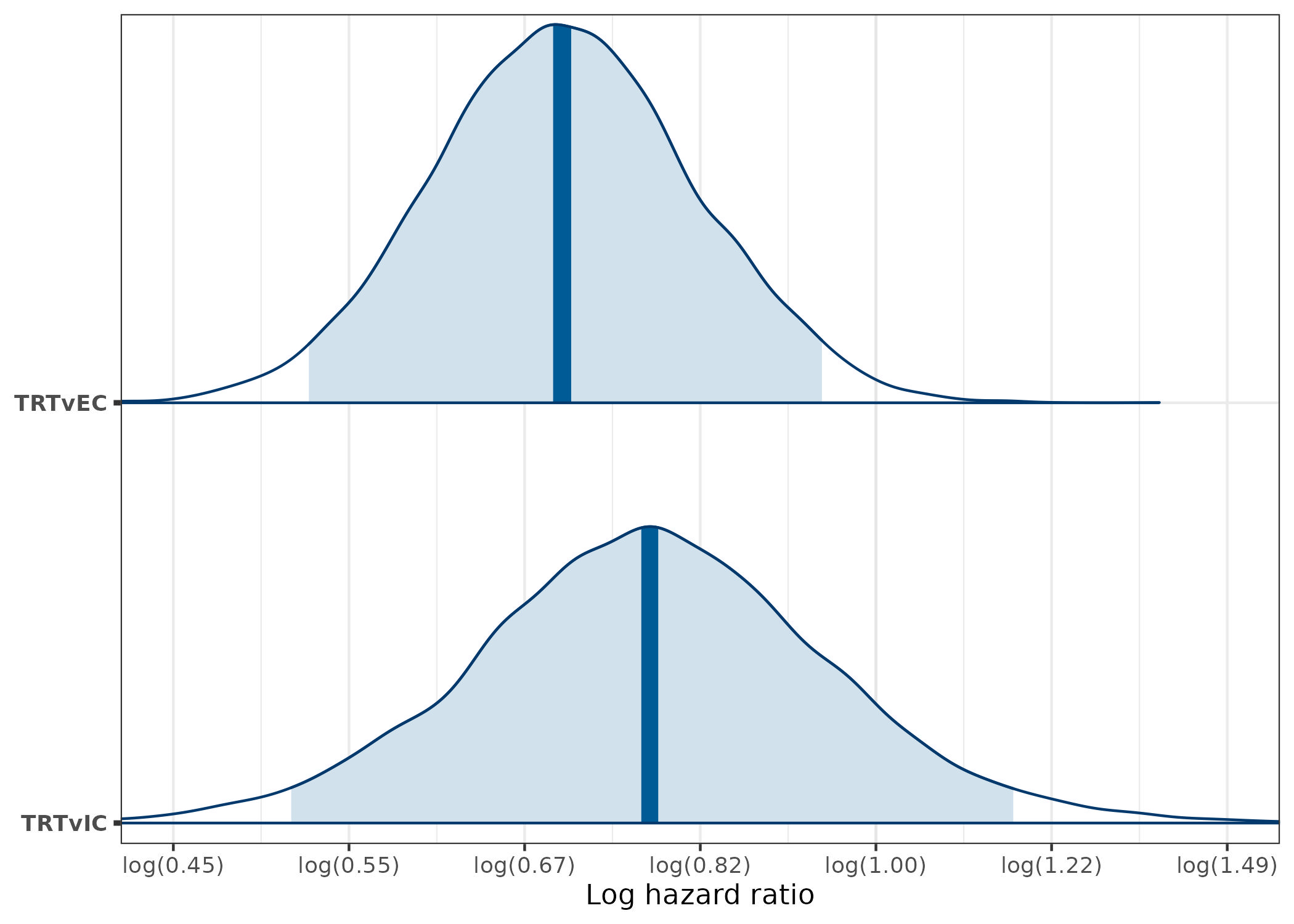} 
\caption{Posterior distributions for log hazard ratios. The top figure is the distribution of the log hazard ratio, $\lambda_{TRTvEC}$, for a comparison of the experimental treatment arm to the external control arm (\emph{TRTvEC}). The bottom figure is the distribution of the log hazard ratio, $\lambda_{TRTvIC}$, for a comparison between the treatment arm and a hypothetical internal control arm (\emph{TRTvIC}), generated after adjusting $\lambda_{TRTvEC}$ using estimates from the meta-analysis assessing the similarity of the internal and external control arms ($\lambda_{ICvEC}$). Shaded areas represent 95\% credible intervals (2.5th percentile to 97.5th percentile) and the vertical line is the posterior median.}
\label{fig:loghr-new-hist}
\end{figure}

To adjust the log HRs, we used a meta-analysis of the estimates from the previous section to generate the posterior predictive distribution for $\lambda^{new}_{TRTvIC}$ as detailed in \autoref{subsec:prediction}. \autoref{fig:loghr-new-hist} demonstrates the adjustment process.\cite{gabry2019visualization} The top figure is the posterior distribution of the log HR from the treatment and external control comparison, $\lambda^{new}_{TRTvEC}$, which is centered around $0.70$ with dispersion determined by the standard error. The bottom figure is the posterior distribution of the log HR from a comparison between the treatment and a hypothetical internal control (i.e., the log HR after ``adjustment''), $\lambda^{new}_{TRTvIC}$. The estimated bias, $\mu$, shifts the posterior distribution rightward so that it is centered (posterior median) at $\log(\hrNewTrtIc)$ in the bottom figure (i.e., the treatment benefit from the naive estimate in the top figure was overly optimistic). Furthermore, consideration of the between study variability, $\sigma$, results in a posterior distribution that is more spread out, meaning that uncertainty intervals for $\lambda^{new}_{TRTvIC}$ are wider than those for $\lambda^{new}_{TRTvEC}$. In this illustration the adjustment was important and could have changed a decision: the log HR was nominally significant at the 5\% level in the naive analysis but was not significant after adjustment.

\subsection{Model checking}
To check the model, we split the reference studies into training and test sets. The model was ``trained" by performing the meta-analysis on the training reference studies and ``tested''  by predicting a treatment effect for the test study, assuming it was a single-arm trial without an internal control. Leave-one-out cross-validation was used to generalize performance, whereby reference studies were partitioned into training and test sets 14 times (with 13 of the 14 reference studies used for training and the remaining reference study used for testing), resulting in 14 separate predictions. For sake of comparison, we made both adjusted and unadjusted predictions. The adjusted predictions, denoted as $\hat{y}$, were given by the posterior median of  $ \lambda^{test}_{TRTvIC}$ and the unadusted predictions were the estimated log HRs from comparisons between the treatment and external control, $\hat{\lambda}^{test}_{TRTvEC}$.

\begin{figure}[h]
\centering
\includegraphics[max size={.9\textwidth}{\textheight}]{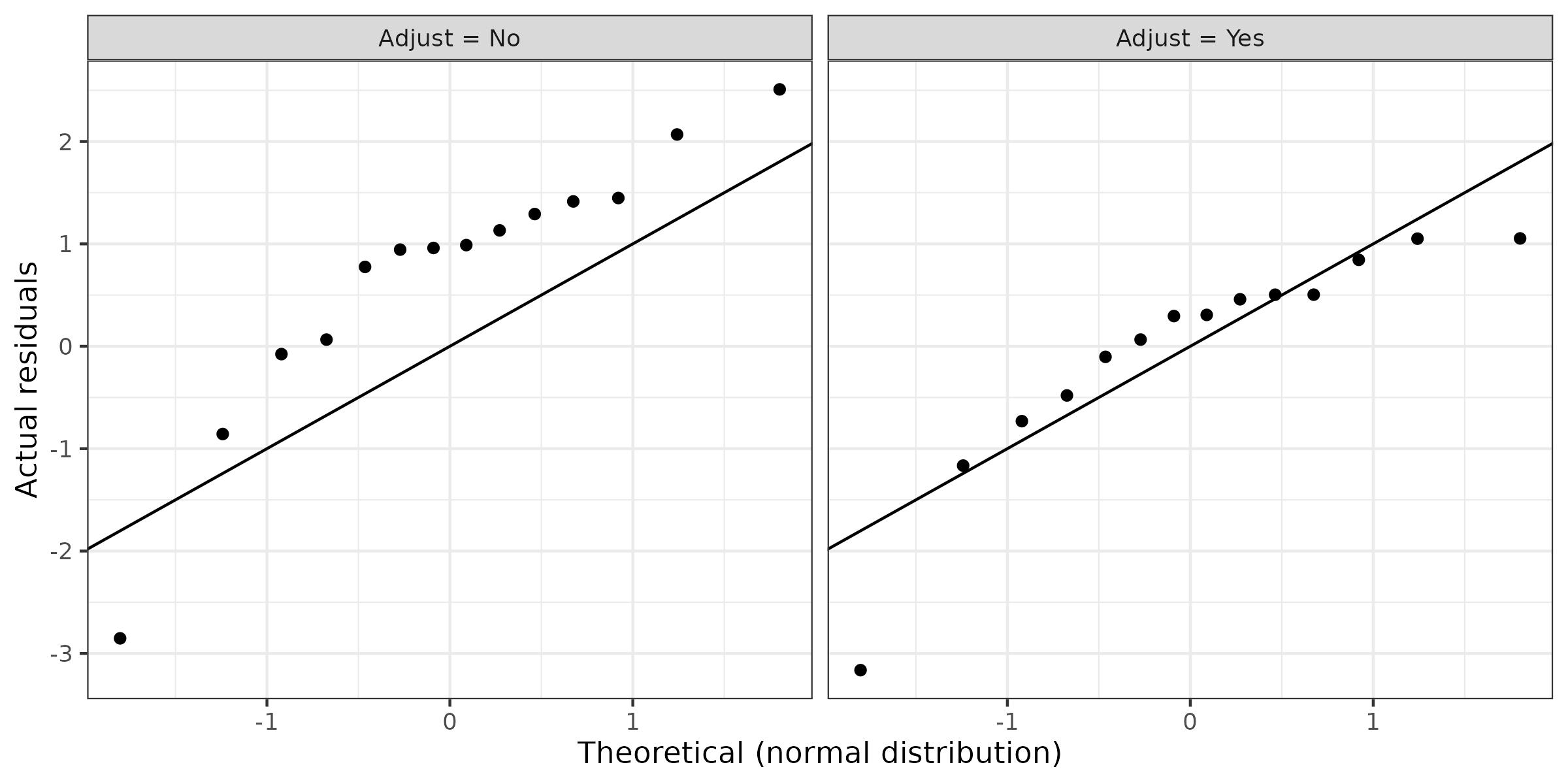} 
\caption{QQ-plot assessing differences between observed and predicted log hazard ratios. The observed log hazard ratios are the estimates from the trial, $\hat{\lambda}^{test}_{TRTvIC}$. The adjusted predictions are given by the posterior median of  $ \lambda^{test}_{TRTvIC}$  and the unadjusted prediction is $\hat{\lambda}^{test}_{TRTvEC}$. The residuals (observed minus predicted) on the y-axis are standardized by dividing by the standard deviation of $ \lambda^{test}_{TRTvIC}$. }
\label{fig:qq-plot}
\end{figure}

To assess the quality of the predictions, we computed residuals, $y - \hat{y}$, where $y$ was the actual log HR estimated from the trial, $\hat{\lambda}^{test}_{TRTvIC}$. Residuals were then standardized by dividing by the standard deviation of $ \lambda^{test}_{TRTvIC}$. A QQ-plot of the standardized residuals against a theoretical normal distribution is displayed in \autoref{fig:qq-plot}.

As expected, the majority of residuals from the unadjusted predictions (left panel) are above the 45 degree line; in other words, the distribution of residuals was centered to the right of $0$ and predictions were therefore biased in favor of the treatment arms. In contrast, predictions in the adjusted case (right panel) are close to the 45 degree line and are residuals are roughly centered around 0, suggesting that the methodology is able to reasonably correct for bias. Reference study 5 is again an outlier as it lies below the 45 degree line in both cases since predicted log HRs were far too high for this study.

\section{Discussion} \label{sec:discussion}
We present a new meta-analytic approach that leverages historical studies to adjust treatment effects in single-arm studies augmented by external controls for the additional bias and variability caused by their non-randomized design. Our example analysis in aNSCLC suggests that this additional bias and variability is important to adjust for and that propensity score methods alone may be unable to completely eliminate confounding.  Across 14 studies, the posterior median of the HR for a comparison between the internal and external controls (after inverse probability weighting) was $\muExpMedian$; that is, survival times were, on average, shorter among external control patients, which would result in a bias in favor of the experimental treatment. Similarly, the between study variability (on the log HR scale) was significant (posterior median: $\sigmaMedian$), suggesting that log HRs vary from study to study beyond what would be expected from sampling error alone. 

Our approach was inspired by the surrogate marker literature that uses meta-analysis to estimate the association between treatment effects for a surrogate marker and treatment effects for a primary clinical outcome.\cite{daniels1997meta, burzykowski2006evaluation}. In this context, the treatment effect from a single-arm study, $\lambda_{TRTvEC}$, is akin to a treatment effect for a surrogate marker and the treatment effect from an RCT, $\lambda_{TRTvIC}$ is analogous to the treatment effect for the primary clinical outcome. An important difference that we leverage in this paper though (\autoref{eqn:hr-relationship}) is that there is a mechanical relationship between $\lambda_{TRTvEC}$ and $\lambda_{TRTvIC}$ determined by $\lambda_{ICvEC}$ . Namely,  $\lambda_{TRTvEC}$ and $\lambda_{TRTvIC}$ both contain the same ``treated'' subjects, which is not the case when assessing the suitability of a surrogate marker. 

While our approach is appealing because it can adjust for unmeasured confounding, there are a number of challenges and questions that must be considered in practice. For one, it is not obvious how the reference studies should be selected or how similar they must be to the single-arm trial of interest. There are a large number of selection criteria that could be considered including the phase of the trial, characteristics of the trial population, drug indication or class, and line of therapy, among others. In some cases a single-arm study may be similar to candidate reference studies on some criteria but not others and analysts will need to determine whether a reference study can be reasonably generalized to the single-arm study. Further study spanning multiple indications and drug classes may be valuable to assess whether the relationships between internal and external controls are generalizable---potentially allowing for extrapolation---or whether they vary across setting and scenario-specific relationships need to be established.

A potential challenge is that there may not be a large set of suitable reference studies, especially for rare diseases or especially novel treatments. Our simulations provide some guidance on the required number of studies and suggest that reasonable estimates can be obtained even when the number of reference studies is smaller. For instance, in cases where there were as few as 4 reference studies, estimates were unbiased and coverage probabilities were too high, implying that conclusions would be conservative. Put differently, our approach helps control type I error even when the number of reference studies is small because it adjusts for bias and results in estimates of the standard error that are conservative. Of course, increasing the number of reference studies is beneficial since it reduces the uncertainty of the estimates and increases power. 

A methodological consideration is the prior chosen for the parameter $\sigma$, representing between study variability. Indeed, when the sample size is small, the scale hyperparameter in a hierarchical model can be sensitive to its prior. In our aNSCLC example, we used a half-Cauchy prior as recommended by Gelman.\cite{gelman2006prior} To assess the sensitivity of this choice, we also considered a gamma prior (scale = 0.001, rate = 0.001) for $1/\sigma^2$ and a uniform prior for $\sigma$ with a lower limit of 0 and a large upper limit of 100. As shown in \autoref{fig:prior-sensitivity}, the posterior distributions for $\sigma$ are very similar when using a uniform or half-Cauchy distribution; however, as noted by Gelman, the gamma prior on the precision results in a posterior distribution that is peaked closer to 0 because the marginal likelihood is higher near 0. A similar pattern was observed in our simulations. For instance, coverages of nominal 95\% confidence intervals were very similar when using half-Cauchy or uniform priors, whereas coverage probabilities were lower when using a gamma prior (\autoref{fig:coverage-comparison}). Likewise, \autoref{fig:power-comparison} shows that power was similar when using a uniform or half-Cauchy prior, but slightly higher with a gamma prior.

A related choice is whether to use a Bayesian or frequentist approach to estimate the meta-analytic model. In some cases, Bayesian methods may be required, particularly if the number of reference studies is too small and an informative prior is needed to estimate $\sigma$. Furthermore, our simulation results (\autoref{fig:coverage-comparison}) suggest that our frequentist approach is too optimistic and results in undercoverage; conversely, the Bayesian approach---especially with a half-Cauchy or uniform prior on $\sigma$---is conservative and results in overcoverage. One potential advantage of the frequentist approach is that it does not take as long to fit, which can be useful in simulation studies; however, in practice, we have found that the Bayesian model converges quickly and that sampling is fast. Taken together, we would generally recommend using a Bayesian approach.

One assumption that is worth revisiting is the relationship between HRs described in \autoref{eqn:hr-relationship}, which will only hold exactly under proportional hazards. We believe this is a reasonable assumption in most clinical trials in oncology with the HR as the primary endpoint and may not be a large concern in practice. The aNSCLC example is a case in point as our methodology seemed to work well. Moreover, in cases where hazards are clearly not proportional, it is unclear whether the HR from a typical unweighted Cox regression model is the best way to measure treatment effects or whether it is even a meaningful quantity.\cite{schemper2009estimation, lin2020alternative, stensrud2020test} In such cases, it would likely be necessary to extend the meta-analysis  to consider differences in survival curves between the internal and external controls, rather than just differences in the log HRs. An alternative would therefore be to consider a multi-dimensional treatment effect, such as the parameters of a parametric \cite{ouwens2010network} or flexible parametric \cite{jansen2011network} survival distribution. 

While the meta-analysis is straightforward to perform, estimation of the log HRs, $\hat{\lambda}_{ICvEC}$, requires a considerable amount of effort, since a separate causal inference analysis must be performed for each reference study. Our approach will consequently make it more burdensome to create a prespecified analysis plan for a single-arm study. In our aNSCLC example, we needed to specify the imputation model for missing data, covariates to control for, functional form of the propensity score model, and propensity score methods (e.g., weighting, matching, etc). We opted to apply a relatively uniform methodology across the 14 studies, but performing a separate analysis for each reference study increases the complexity of the analysis and programming.  

A more general question relevant to both the new single-arm study and the meta-analysis is whether propensity score methods should be fully prespecified or if some discretion should be left to the analyst. One advantage of propensity score methods is that the design of the study can be separated from the analysis so that the propensity score method can be refined without knowledge of the outcomes.\cite{rubin2001using, rubin2008objective, stuart2010matching} It may, for instance, be reasonable to tweak the specification of the propensity score model after assessing its impact on the balance and overlap of baseline covariates.\cite{rosenbaum1984reducing, austin2008critical,belitser2011measuring}. We chose to do this in our analysis and found it useful as certain specifications such as inclusion of interaction terms resulted in worse balance and more extreme weights. However for use in a regulatory setting pre-specification may be required or at a minimum demonstration that any changes were made without any knowledge of the outcomes. Otherwise, there is considerable risk that estimates of bias will be underestimated.

When creating the set of reference studies, an additional question is whether specification of the propensity score model used to ultimately compare the internal and external controls should be refined using the combined trial data, the experimental arm alone, or the internal control arm alone. An argument in favor of using the internal control arm alone is that the meta-analysis does not require the experimental arm. Yet, we decided to assess the propensity score model based on an analysis that only included the experimental trial arm, since the internal control arm would not be available in a single-arm trial. That said, in practice, these arms should have similar characteristics because they are randomized and the decision should have little impact on the conclusions. 

Our meta-analytic approach may also be useful outside of the prospective single-arm setting. Due to the increasing interest in the use of RWD for drug development, there is both a need and an interest in assessing the extent to which RWD can emulate arms from completed clinical trials. Current emulation efforts include comparative-effectiveness studies \cite{franklin2021emulating} in which RWD is used to construct both the treatment and control arms and external control studies with a trial experimental arm and RWD control arm \cite{carrigan2020using}. However, it is unclear how to determine whether an emulation was successful. In the external control setting, the meta-analysis described in this paper comparing the internal and external control arms provides one such validation tool. The advantage of a formal meta-analysis is that it estimates both bias and between study variability as well the uncertainty of those estimates, while accounting for the sampling error in each of the validation studies. A similar approach based on the surrogate marker literature could be used for comparative effectiveness studies, with the RCT treatment effect serving as the primary endpoint and the RWD (emulated study) treatment effect as the surrogate marker.

\section{Conclusion} \label{sec:conclusion}
External control arms can be used to estimate treatment efficacy in single-arm trials, but bias and potential unlimited inflation of type 1 error is a serious concern because the external control arm is not randomized. To address this problem, we have proposed a meta-analytic framework that uses a set of historical reference studies to adjust estimates in a new single-arm study for the additional bias and variability caused by the lack of randomization. The approach is straightforward to apply since existing causal inference methods typically used in observational studies do not need to be modified in any way and predictions in a new study can be implemented using our software. By minimizing the risk of false positives, we believe that our approach can help increase acceptance of external control studies in both internal---such as ``go/no-go'' decisions on whether to move a molecule forward for further study---and regulatory decision making.

\bibliographystyle{ama}
\bibliography{references}

\section*{Data availability statement}
Data from Flatiron Health, Inc were used to construct the external control cohorts. These de-identified data may be made available upon request and are subject to a license agreement with Flatiron Health; interested researchers should contact Flatiron Health to determine licensing terms. For the internal control cohorts for clinical trials 1, 2, 3, 12 and 13 qualified researchers may request access to individual patient level clinical data for each separate study through a data request platform. At the time of writing this request platform is Vivli (\url{https://vivli.org/ourmember/roche/}). For the remaining clinical trials qualified researchers may submit an enquiry to Vivli to request access to individual patient level clinical data. For up to date details on Roche's Global Policy on the Sharing of Clinical Information and how to request access to related clinical study documents, see here:  \url{https://go.roche.com/data_sharing}.

The code used to produce the results reported in the paper is available at \url{https://github.com/phcanalytics/ecmeta-manuscript}. Documentation and installation instructions for the \textsf{R} package \pkg{ecmeta} can be found at \url{https://github.com/phcanalytics/ecmeta}.

\appendix
\renewcommand\thefigure{\thesection.\arabic{figure}}    
\setcounter{figure}{0} 

\section{Maximum likelihood approach} \label{appendix:ml}
\subsection{Estimation}
$\mu$ and $\sigma$ can be estimated by optimization of the log-likelihood function,

\begin{equation}
\text{log } l(\mu,\sigma) = 0.5 \times \sum{log(\sigma^2+V_j)} +
0.5 \times \sum{ \frac{(\hat\lambda_j-\mu)^2}{(\sigma^2+V_j)} }.
\end{equation}

\subsection{Prediction for a new study}
Following Gelman et al. (see p. 66) \cite{gelman2013bayesian}, we note that the posterior predictive distribution for a new observation $y^{new}$ follows a t distribution with location $\overline{y}$, scale $s\sqrt{(1 + 1/n)}$, and $n-1$ degrees of freedom where $\overline{y}$ is the mean and $s^2=1/(n-1)\sum (y_i - \overline{y})^2$ is the variance from the $n$ observations. In our problem it follows that $\lambda_{ICvEC}^{new}$ follows a t distribution with mean $\hat{\mu}$ and scale $\hat{\sigma}\sqrt{(1 + 1/n)}$.

A simulation procedure for obtaining draws of $\lambda_{TRTvIC}^{new}$ then proceeds as follows:

\begin{enumerate}
\item Draw $\lambda_{TRTvEC}^{new} \sim N(\hat{\lambda}_{TRTvEC}^{new}, V^{new})$ 

\item Draw $\lambda_{ICvEC}^{new} \sim t_{n-1}(\hat{\mu}, \hat{\sigma}\sqrt{(1 + 1/n)})$ where $t_v(a, b)$ is a student t distribution with $v$ degrees of freedom, location $a$, and scale $b$

\item Use the draws from 1 and 2 to compute $\lambda_{TRTvIC}^{new} =  \lambda_{TRTvEC}^{new} - \lambda_{ICvEC}^{new}$
\end{enumerate}

\section{Propensity score methodology} \label{appendix:ps}
This section expands on the propensity score methodology from \autoref{subsec:propensity-score} and describes the sensitivity analyses. All results and code are available online in the GitHub repository for the paper.

\subsection{Coding of covariates}
Although we aimed to code categorical covariates in a consistent way across the reference studies, this was not always possible. Race was typically coded as White versus Other since the number of observations within non-White race categories was usually very small. Two exceptions were NCT02367781 and NCT01351415. We used three categories (White, Black, Other) for the former since there were a significant number of Black patients in the RCT; for the latter, race was coded as Asian or Non-Asian since it was coded in that way in the trial. Similarly, histology was only included if it was not part of the inclusion and exclusion criteria for a trial and covariates were excluded if they were not collected for a particular trial.

We explored both linear terms and restricted cubic splines with 4 knots placed at equally spaced quantiles of the support for the continuous variables (age and time since initial diagnosis). In addition, an interaction between cancer stage and time since diagnosis was considered based on the models employed in Carrigan et al.\cite{carrigan2020using} A specification was chosen after evaluation of the balance of the covariates subsequent to population adjustment with the propensity score methods. We ultimately chose the model with linear terms and no interactions since it tended to  result in the best balance and was less likely to generate extreme weights. 

\subsection{Sensitivity analyses}

\subsubsection{Trimming}
Recall that the primary analysis used IPTW-ATT weights and removed external control patients with propensity scores below the 1st percentile and above the 99th percentile. A sensitivity analysis was also performed in which no patients were removed from the sample.

\subsubsection{Propensity score matching}
Propensity score matching was used as a sensitivity analysis. Two approaches were utilized. First, we matched patients on the linear propensity score with greedy 1:1 nearest neighbor matching. For a given treated subject, 1:1 nearest neighbor matching selects the control subject whose value of the linear propensity score is closest; that is, for a treated subject $i$, the control subject $j$ with the minimum value of $D_{ij}=|\rm{logit}(e_i)-\rm{logit}(e_j)|$ was chosen where $e_{k}$ is the propensity score for subject $k$. In cases where multiple control subjects were equally close to a treated subject, a treated subject was chosen at random. We believe 1:1 matching is reasonable given simulation evidence showing that mean square error is typically minimized when matching either 1 or 2 controls to each treated subject.\cite{austin2010statistical} Greedy matching meant that control subjects were chosen one at a time and the order in which they were chosen mattered. Matching was performed without replacement (except in cases where the number of control subjects was less than the number of treated subjects) to ensure that the matched controls were independent, although it is worth noting that there is some evidence that matching with replacement can reduce bias.\cite{abadie2006large, stuart2010matching}.

Our second approach used genetic matching, which can model the propensity score in a more flexible way.\cite{sekhon2008multivariate, diamond2013genetic}  In particular, one challenge of nearest neighbor matching---and propensity core methods more generally---is that it is difficult to specify a model that achieves satisfactory covariate balance. Genetic matching can help since the weight given to each covariate is based on an evolutionary search algorithm that iteratively checks and improves balance. We implemented this algorithm using both the linear propensity score and all the covariates from the propensity score model, so that matching on the propensity score and the Mahalanobis distance were limiting cases.

Both matching algorithms  were performed with and without a caliper. In the case of the former, a caliper on the linear propensity score of 0.25 standard deviation was used.\cite{rosenbaum1985constructing}.

\subsubsection{Double adjustment}
To facilitate simple estimation of marginal HRs, the Cox models in the primary analysis did not adjust for covariates. A sensitivity analysis was performed in which regression adjustment was used to estimate hazard ratios (i.e., by including all covariates used in the propensity score model in the Cox model). \cite{rubin1973use, nguyen2017double}.

\section{Supplementary figures} \label{appendix:figs}

\begin{figure}[h]
\centering
\includegraphics[max size={.95\textwidth}{\textheight}]{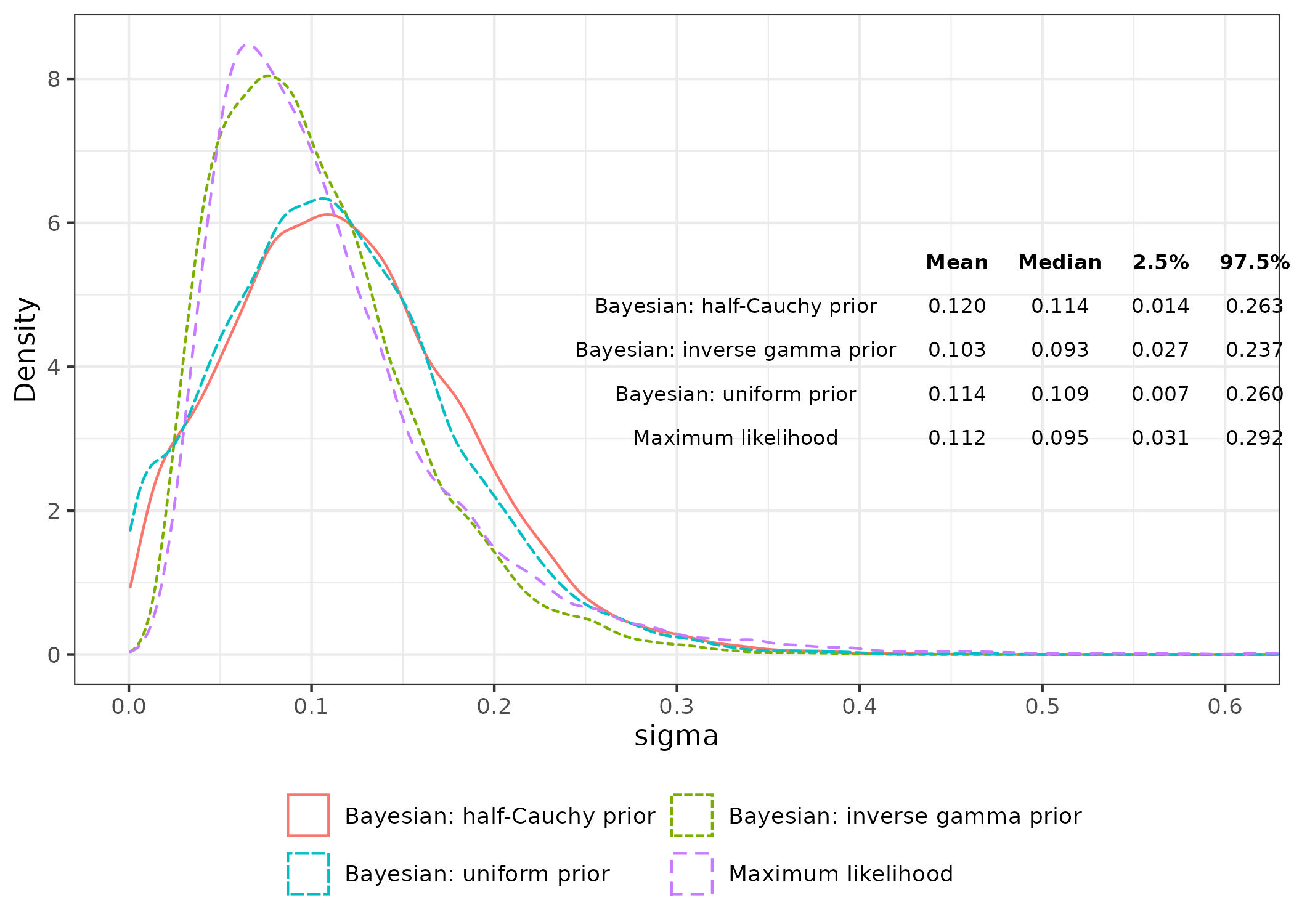} 
\caption{Distribution of the between study variability parameter, $\sigma$, by analysis method. The half-t and uniform priors are on $\sigma$ and the inverse gamma prior is on $\sigma^2$. The half-Cauchy prior has location  0 and scale 25; the inverse gamma prior has shape 0.001 and rate 0.001; and the uniform prior has a lower limit of 0 and upper limit of 100. }
\label{fig:prior-sensitivity}
\end{figure}

\begin{figure}[h]
\centering
\includegraphics[max size={.95\textwidth}{\textheight}]{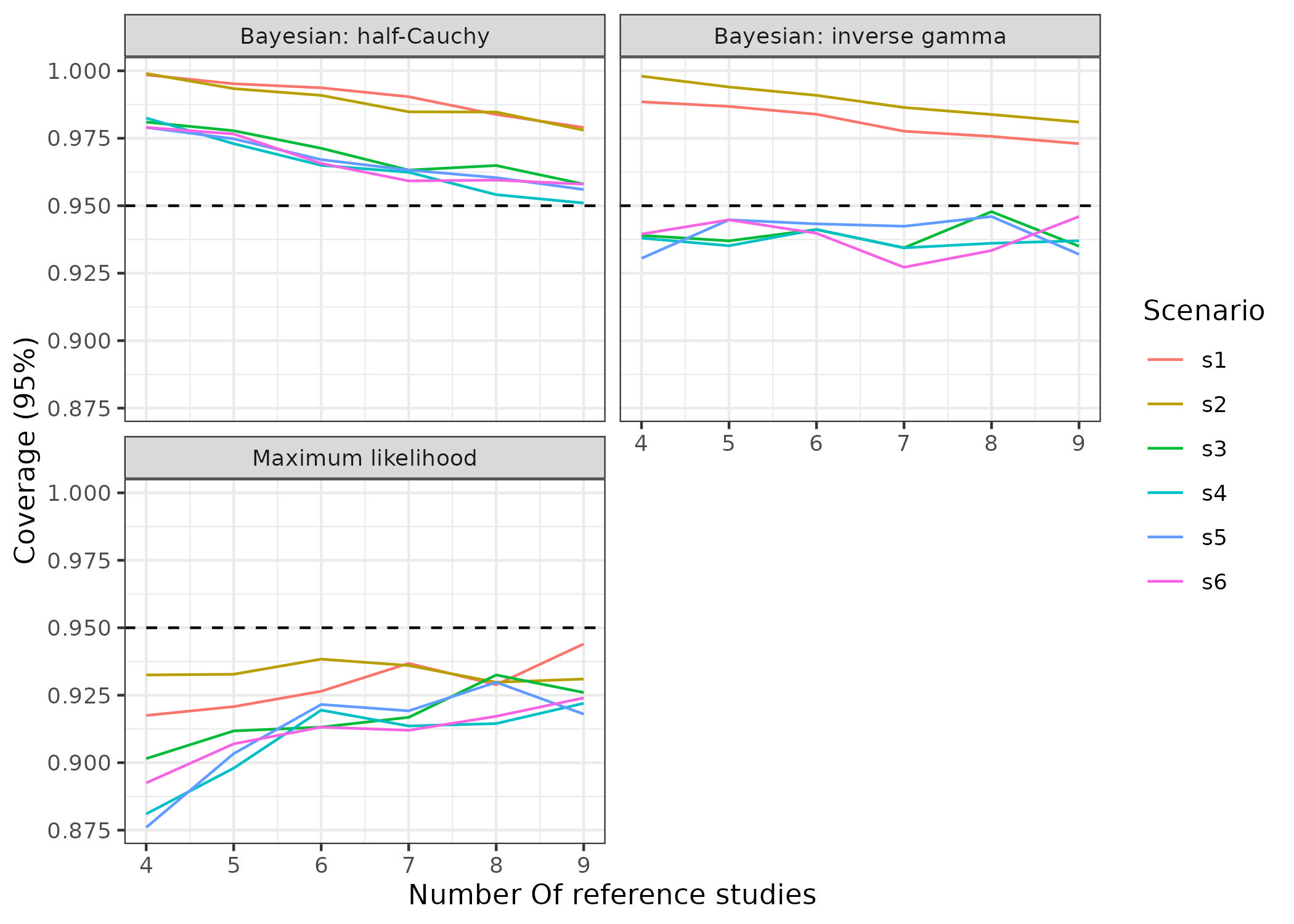} 
\caption{Coverage of 95\% confidence intervals by analysis method, scenario, and the number of reference studies.}
\label{fig:coverage-comparison}
\end{figure}

\begin{figure}[h]
\centering
\includegraphics[max size={.95\textwidth}{\textheight}]{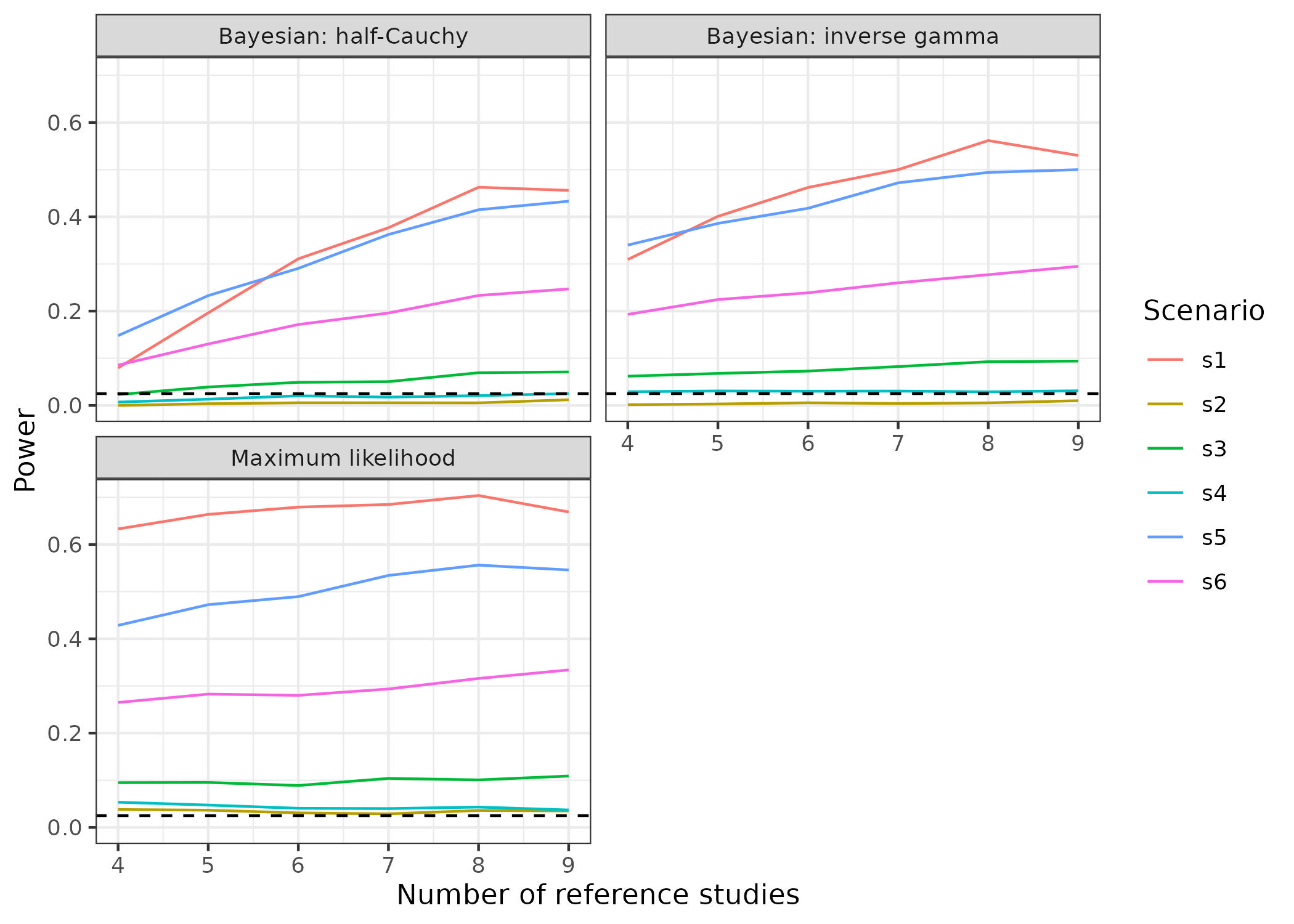} 
\caption{Power by analysis method, scenario, and the number of reference studies.}
\label{fig:power-comparison}
\end{figure}

\end{document}